\documentclass[pdflatex,sn-mathphys-num]{sn-jnl}

\usepackage{graphicx}%
\usepackage{multirow}%
\usepackage{amsmath,amssymb,amsfonts}%
\usepackage{amsthm}%
\usepackage{mathrsfs}%
\usepackage[title]{appendix}%
\usepackage{xcolor}%
\usepackage{textcomp}%
\usepackage{manyfoot}%
\usepackage{booktabs}%
\usepackage{algorithm}%
\usepackage{algorithmicx}%
\usepackage{algpseudocode}%
\usepackage{listings}%
\usepackage{braket}

\begin{document}
	
	\title[Article Title]{Photon acceleration of high-intensity vector vortex beams into the extreme ultraviolet}
	
	
	\author*[1]{\fnm{Kyle G.} \sur{Miller}}\email{kmill@lle.rochester.edu}
	\author[2]{\fnm{Jacob R.} \sur{Pierce}}\email{jacobpierce@physics.ucla.edu}
	\author[2]{\fnm{Fei} \sur{Li}}\email{lifei11@g.ucla.edu}
	\author[3]{\fnm{Brandon K.} \sur{Russell}}\email{bkruss@umich.edu}
	\author[2,4]{\fnm{Warren B.} \sur{Mori}}\email{mori@physics.ucla.edu}
	\author[3]{\fnm{Alexander G. R.} \sur{Thomas}}\email{agrt@umich.edu}
	\author[1]{\fnm{John P.} \sur{Palastro}}\email{jpal@lle.rochester.edu}
	
	\affil*[1]{\orgdiv{Laboratory for Laser Energetics}, \orgname{University of Rochester}, \orgaddress{\city{Rochester}, \postcode{14623-1299}, \state{NY}, \country{USA}}}
	\affil[2]{\orgdiv{Department of Physics and Astronomy}, \orgname{University of California, Los Angeles}, \orgaddress{\city{Los Angeles}, \postcode{90095}, \state{CA}, \country{USA}}}
	\affil[3]{\orgdiv{G\'erard Mourou Center for Ultrafast Optical Sciences}, \orgname{University of Michigan}, \orgaddress{\city{Ann Arbor}, \postcode{48109}, \state{MI}, \country{USA}}}
	\affil[4]{\orgdiv{Department of Electrical and Computer Engineering}, \orgname{University of California, Los Angeles}, \orgaddress{\city{Los Angeles}, \postcode{90095}, \state{CA}, \country{USA}}}
	
	\abstract{
		Extreme ultraviolet (XUV) light sources allow for the probing of bound electron dynamics on attosecond scales, interrogation of high-energy-density matter, and access to novel regimes of strong-field quantum electrodynamics. Despite the importance of these applications, coherent XUV sources remain relatively rare, and those that do exist are limited in their peak intensity and spatio-polarization structure. Here, we demonstrate that photon acceleration of an optical vector vortex pulse in the moving density gradient of an electron beam--driven plasma wave can produce a high-intensity, tunable-wavelength XUV pulse with the same vector vortex structure as the original pulse. Quasi-3D, boosted-frame particle-in-cell simulations show the transition of optical vector vortex pulses with 800-nm wavelengths and intensities below $10^{18}$~W/cm$^2$ to XUV vector vortex pulses with 36-nm wavelengths and intensities exceeding $10^{20}$~W/cm$^2$ over a distance of 1.2~cm. The XUV pulses have sub-femtosecond durations and nearly flat phase fronts. The production of such high-quality, high-intensity XUV vector vortex pulses could expand the utility of XUV light as a diagnostic and driver of novel light--matter interactions.}
	
	
	\maketitle
	
\begin{figure*}[h]
		\centering
		\includegraphics[width=0.49\textwidth]{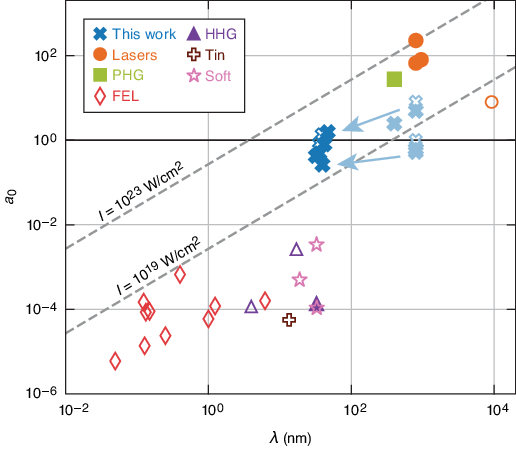}
		\caption{Amplitude of available sources from the XUV to the long-wave infrared. Here $a_0 = 8.5\times10^{-10} [I(\mathrm{W/cm}^2)]^{1/2} \lambda(\mu \mathrm{m})$ is the amplitude threshold for relativistic electron motion and $I$ is intensity. A sizable gap in amplitude and intensity exists between available XUV and near-optical sources. This gap exists for both conventional pulses (open symbols) and those with spatio-polarization structure (solid symbols). Photon acceleration of structured optical pulses in an electron beam--driven plasma wave offers a tunable-wavelength source that can bridge this gap (span of light to dark blue solid X symbols). Currently available sources include conventional lasers (Lasers), perturbative harmonic generation (PHG)~\cite{Wang20170.85Beamline}, free-electron lasers (FELs)~\cite{Gutt2012SingleRegime,Chollet2015TheSource,Matsuyama2016NearlyMirrors,Decking2020AAccelerator,Eom2022RecentPAL-XFEL,Grychtol2022TheXFEL}, high harmonic generation (HHG)~\cite{Dromey2006HighLimit,delasHeras2022Extreme-ultravioletGeneration}, in-band emission from laser-produced tin plasmas (Tin)~\cite{Versolato2022Microdroplet-tinReview}, and collisional soft x-ray lasers (Soft)~\cite{Zeitoun2004ABeam,Wang2008Phase-coherent13.9nm,Depresseux2015Table-topGating}.}\label{fig:sources}
	\end{figure*}

	Extreme ultraviolet (XUV) light sources facilitate technological advances and scientific discoveries across multiple disciplines. XUV pulses are used to image ``molecular movies''~\cite{Glownia2016Self-ReferencedMotion,Coquelle2018ChromophoreCrystallography,ZhouHagstrom2022Megahertz-rateXFEL}, interrogate high-energy-density and warm dense matter~\cite{Glenzer2009X-rayPlasmas,Levy2009X-rayMatter,Fletcher2015UltrabrightPhysics}, perform nanometer-scale photolithography and imaging of microprocessors~\cite{Sakdinawat2010NanoscaleImaging,Maldonado2016X-rayProspects,Holler2019Three-dimensionalZoom}, and resolve attosecond dynamics of bound electrons~\cite{Hockett2011Time-resolvedReaction}. The importance of these applications has, in turn, fueled the development of XUV sources, including free-electron lasers~\cite{Gutt2012SingleRegime,Chollet2015TheSource,Matsuyama2016NearlyMirrors,Decking2020AAccelerator,Eom2022RecentPAL-XFEL,Grychtol2022TheXFEL}, high harmonic generation in solids and gases~\cite{Dromey2006HighLimit,delasHeras2022Extreme-ultravioletGeneration}, in-band emission from laser-produced tin plasmas~\cite{Versolato2022Microdroplet-tinReview}, and collisional soft x-ray lasers~\cite{Zeitoun2004ABeam,Wang2008Phase-coherent13.9nm,Depresseux2015Table-topGating}. Despite ongoing and rapid progress in the development of these sources, a sizable gap exists between the intensities available at optical frequencies and those available in the XUV (Fig.~\ref{fig:sources}). Bridging this intensity gap would improve the signal-to-noise ratio in critical XUV diagnostics and enable novel regimes of strong-field ionization~\cite{Popruzhenko2009MultiphotonLasers,Kanter2011UnveilingPulses} and quantum electrodynamics~\cite{Blaschke2009DynamicalQED,DiPiazza2012ExtremelySystems}.
	
	Alongside their ability to create high intensities, optical sources allow for the use of conventional, adaptive, or metasurface optics to manipulate the spatial, temporal, or polarization structure of the pulses they produce~\cite{Aieta2015MultiwavelengthCompensation,Kondakci2019OpticalSpace,Shaltout2019SpatiotemporalMetasurfaces,Ambat2023Programmable-trajectoryPulses}. These optics have the advantage of linearity, which helps ensure that the manipulation is controlled and deterministic. This contrasts the inherently nonlinear processes that have been used to generate structured XUV pulses~\cite{delasHeras2022Extreme-ultravioletGeneration}. Within the field of structured light, vector vortex beams~\cite{Padgett2004LightsMomentum,Liu2017GenerationSphere}---which generalize and couple the spatial, orbital-angular-momentum, and polarization degrees of freedom---have found utility in optical trapping~\cite{Padgett2011TweezersTwist,Yang2021OpticalReview}, spectroscopy~\cite{Sirenko2019TerahertzExcitations}, quantum information~\cite{Wang2012TerabitMultiplexing,Erhard2017TwistedDimensions,Ndagano2017CharacterizingLight}, particle acceleration~\cite{Marceau2013FemtosecondGas}, and material processing~\cite{Meier2007MaterialRadiation}. Overcoming the ``structure gap'' between optical and XUV pulses with more-controlled techniques for creating vector vortex XUV pulses would expand their flexibility as diagnostics and drivers of novel light--matter interactions. 
	
	Photon acceleration~\cite{Wilks1989PhotonAccelerator,Dias1997ExperimentalFronts,Mori1997,Mendonca2000TheoryAcceleration,Murphy2006EvidenceFields,Shcherbakov2019PhotonMetasurfaces,Howard2019PhotonFocus,Franke2021OpticalAcceleration} has the potential to convert high-intensity optical pulses to ultrafast, high-intensity XUV pulses while preserving the initial spatio-polarization structure. Photon acceleration refers to the temporal refraction of photons in a time-dependent refractive index.  Just as a spatially varying refractive index changes the propagation direction of light rays, a time-varying refractive index changes the frequency of light. In a recently proposed photon acceleration scheme, a linearly polarized laser pulse co-propagates with and frequency upshifts in the density modulation of a plasma wave driven by a relativistic electron beam~\cite{Sandberg2023PhotonXUV,Sandberg2023Electron-beam-drivenPhotons,Sandberg2024DephasinglessAcceleration}. This scheme has two beneficial properties that offer a path to surmounting the intensity and structure gaps: First, a high-intensity laser pulse can extract energy from the plasma wave and maintain its amplitude as it upshifts, resulting in a high-intensity XUV pulse. Second, the plasma wave provides a guiding structure that can prevent diffraction and preserve the spatio-polarization structure of the high-intensity pulse.
	
	Here, we demonstrate the production of near-diffraction-limited, high-intensity, attosecond XUV pulses with predetermined spatio-polarization structure. Quasi-3D particle-in-cell (PIC) simulations show that the plasma wave driven by a 50-GeV, 5.8-nC electron beam can guide and photon accelerate a relativistically intense, vector vortex pulse from a wavelength of 800~nm to 36~nm over 1.2~cm. The initial pulse corresponds to the output of a Ti:sapphire laser; the electron beam parameters are similar to those previously achieved at SLAC~\cite{Alley1995TheSource} and were chosen to mitigate beam evolution over the length of the accelerator. The optical wavelength of the pulse allows for preliminary structuring with linear optical elements to form the vector vortex profile. This profile is preserved, the duration is compressed to attoseconds, and the intensity is enhanced by more than two orders of magnitude during the acceleration. To match the velocity of the accelerating pulse to that of the plasma wave, the  background plasma density decreases over the length of the accelerator~\cite{Sandberg2023PhotonXUV}. The PIC simulations were conducted with \textsc{Osiris}~\cite{Fonseca2002a} and made possible by the use of a Lorentz-boosted frame, which significantly reduces the computational cost of resolving the wavelength of the XUV pulse. These results indicate that photon acceleration can bridge the intensity and structure gap between optical and XUV pulses (Fig.~\ref{fig:sources}).
	
	\section*{Results}\label{sec:results}
	
	\subsection*{Photon-accelerated vector vortex beams}\label{sec:simulation}
	In a photon accelerator, a moving refractive index gradient alters the local phase velocity of a light wave. If the light wave co-travels with a negative (positive) refractive index gradient in a normally (anomalously) dispersive medium, the phase fronts compress, causing a continual increase in the frequency and group velocity. The local frequency $\omega$ and wave vector $k$ of the light wave are related through the refractive index $\eta(\omega,\zeta,z) = ck/\omega$, where $z$ is the propagation axis, $\zeta = t - z/v_\mathrm{I}$ is the moving-frame coordinate, and $v_\mathrm{I}$ is the velocity of the refractive index gradient. In the ray description of light, the frequency acts as the Hamiltonian $\omega =ck/\eta(\omega,\zeta,z)$ and thus evolves according to
	\begin{equation}\label{eq:EOM1}
		\frac{\mathrm{d}\omega}{\mathrm{d}t} = -\frac{\omega}{\eta}\frac{\partial}{\partial \zeta}\eta(\omega,\zeta,z),
	\end{equation}
	where $\partial_t = \partial_{\zeta}$ has been used~\cite{Mendonca2000TheoryAcceleration}. Equation~\eqref{eq:EOM1} shows that photon acceleration to high frequencies over short distances requires large and ultrafast gradients in the refractive index ($\partial_{\zeta} \eta$). 
	
	Plasma is composed of highly mobile free electrons that can support large, ultrafast gradients in the refractive index. This same property allows plasma to maintain its dielectric properties in the presence of high-intensity light waves. An unmagnetized plasma exhibits normal dispersion characterized by the refractive index $\eta = (1-n/\gamma n_\mathrm{c})^{1/2}$, where $n$ is the electron density, $n_\mathrm{c} = m_\mathrm{e} \varepsilon_0 \omega^2/e^2$ is the critical density, and $\gamma$ is the electron energy normalized to $m_\mathrm{e}c^2$. Plasma-based photon accelerators exploit this density dependence in the refractive index. Relativistic charged-particle beams or intense laser pulses can drive large-amplitude, moving density modulations $\delta n(\zeta,z) =  n(\zeta,z) - n_0(z)$, where $n_0(z)$ is the upstream, unperturbed plasma density. The moving refractive index gradient associated with these modulations ($\partial_\zeta \eta \propto \partial_\zeta \delta n)$ travels at a velocity that is nearly equal to that of the driver $v_\mathrm{d}$ and a trailing light wave $v_\mathrm{g}$, i.e., $v_\mathrm{I} \approx v_\mathrm{d} \approx v_\mathrm{g}$. Relativistic electron beams, in particular, are resilient to nonlinear modifications and can drive a steady, ultrafast accelerating gradient over the distances required for a trailing laser pulse to reach high frequencies.
	
\begin{figure*}[h]
		\centering
		\includegraphics[width=0.98\textwidth]{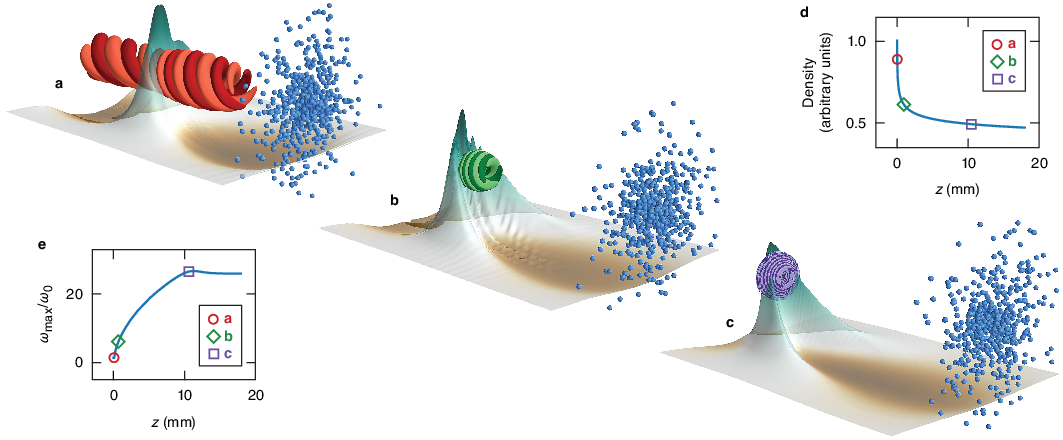}
		\caption{Electron beam--driven photon acceleration of a vector vortex laser pulse. (a)--(c)~An electron beam (blue spheres) traveling through a plasma excites a moving electron density perturbation (brown/blue surface) that photon accelerates a trailing vector vortex pulse (electric field in red, green, and purple). The three frames correspond to sequential locations along the accelerator and are demarcated by the red, green, and purple shapes in (d) and (e). (d)~The tailored density decreases to synchronize the motion of the electron density gradient and the temporal centroid of the vector vortex pulse. (e)~The maximum frequency upshifts by a factor of 27. Here, the vector vortex pulse is a superposition of orthogonally polarized Laguerre--Gaussian modes with azimuthal mode numbers $\ell = 2$ and $-2$ and radial mode number zero. The vector vortex structure is preserved throughout the process. The images were generated from the output of quasi-3D, boosted-frame PIC simulations (see Methods).}\label{fig:schematic}
	\end{figure*}

	A simulation of such a photon accelerator is displayed in Fig.~\ref{fig:schematic}. An ultrashort electron beam and a trailing laser pulse with vector vortex structure~(a)--(c) propagate through background plasma with a ramped density profile~(d), resulting in a frequency upshift of the pulse~(e). The space-charge force of the electron beam expels background electrons from its path. The expelled electrons are then attracted back to the propagation axis by the ions. This results in a nonlinear plasma wave, shown by the brown/blue surface plots of the electron density in (a)--(c). The density modulation of the plasma wave provides the moving refractive index gradient needed to guide and accelerate the trailing vector vortex pulse. The front half of the pulse straddles the rising edge of the density modulation and frequency upshifts due to the larger and smaller phase velocities behind and ahead of the rising edge, respectively. Similarly, the rear half of the pulse straddles the falling edge of the density modulation and experiences a frequency downshift. These frequency shifts are observed on either side of the density spike in (a).
	
	In order for the laser pulse to continually upshift, it must remain overlapped with the rising edge of the density modulation. As the pulse upshifts in frequency, its group velocity $v_\mathrm{g} \approx c\eta$ increases. If the unperturbed plasma density $n_0$ were uniform, the location of the maximum density gradient would travel at the speed of the electron beam $v_\mathrm{d} = v_\mathrm{I} \approx c > v_\mathrm{g}$. Thus, despite accelerating, the pulse would recede with respect to the maximum gradient and eventually slip into the falling edge of the modulation and begin downshifting. To ensure continual upshifting, the unperturbed density profile $n_0(z)$ can be tailored~\cite{Sandberg2023PhotonXUV} so that the maximum density gradient, i.e., where $\delta n = 0$, tracks the temporal centroid of the pulse [Fig.~\ref{fig:schematic}(d)]. The location behind the electron beam where $\delta n = 0$ depends on the wavelength of the plasma wave, which is proportional to $1/\sqrt{n_0}$. If the unperturbed plasma density has a downramp, the plasma wavelength elongates as the electron beam moves into lower densities, causing the maximum gradient to recede. Thus, the profile of $n_0(z)$ can be tailored so that the pulse and maximum gradient recede at the same rate with $v_\mathrm{I}(z) = v_\mathrm{g}(z)$.
	
	Determination of the tailored density profile requires solving a system of two coupled equations. The first equation describes the spatial evolution of the frequency and is obtained from Eq.~\eqref{eq:EOM1}:
	\begin{equation}\label{eq:freq}
		\frac{\mathrm{d}\omega}{\mathrm{d}z} = \frac{e^2}{2\varepsilon_0 m_\mathrm{e}c^2k_z} \frac{\partial(\delta n/\gamma)}{\partial \zeta},
	\end{equation}
	where $k_z$ is the longitudinal wave vector and $\mathrm{d}z/\mathrm{d}t = v_\mathrm{g} = c^2k_z/\omega$ has been used.
	The second equation is derived by setting the location of the temporal centroid $\zeta_\mathrm{c}(z)$ equal to the location where the density modulation of the plasma wave is zero: $\zeta_\delta(z)$ satisfying $\delta n[\zeta_\delta(z),z] = 0$. In differential form, this condition is $\mathrm{d}\zeta_\mathrm{c}/\mathrm{d}z = \mathrm{d}\zeta_\delta/\mathrm{d}z$, or
	\begin{equation}\label{eq:den}
		\frac{1}{v_\mathrm{g}} -\frac{1}{c} = \frac{\mathrm{d}\zeta_\delta}{\mathrm{d}n_0} \frac{\mathrm{d}n_0}{\mathrm{d}z} \quad \Rightarrow \quad \frac{\mathrm{d}n_0}{\mathrm{d}z} \approx \frac{n_0}{2n_\mathrm{c}(\omega)}\left[ \frac{\mathrm{d}\zeta_\delta}{\mathrm{d}n_0}\right]^{-1},
	\end{equation}
	where the approximations $v_\mathrm{d} = c$ and $\gamma(\zeta_\delta)=1$ have been used.
	Closing the system of Eqs.~\eqref{eq:freq} and \eqref{eq:den} requires either analytic expressions or numerical solutions for the density modulation $\delta n$ and location of its zero $\zeta_{\delta}$ \cite{Sandberg2023PhotonXUV}. These can be obtained from a quasistatic model of the electron beam--driven plasma wave (see Methods).
	
	The exact density profile $n_0(z)$ and maximum frequency shift depend on the parameters of the electron beam and trailing laser pulse. The charge of the electron beam should be large enough to create a nonlinear density depression and spike that are wide radially [brown and blue in Figs.~\ref{fig:schematic}(a)--(c)], but not so large that the density spike is too narrow longitudinally. A wide density depression acts like a waveguide that confines the trailing laser pulse transversely and mitigates diffraction as it accelerates. This is particularly important for higher-order vector vortex modes, which have a larger radial extent.
	A slightly diffuse density spike provides a longer accelerating region in $\zeta$ than a narrow spike, which reduces sensitivity to the exact location of the trailing pulse. The energy of the beam must be sufficiently high that it is resilient to the electrostatic fields of the plasma wave over the length of the accelerator. A 50-GeV, 5.8-nC drive beam that roughly corresponds to previous SLAC parameters~\cite{Alley1995TheSource} satisfies these conditions and provides a steady accelerating structure over 1.2~cm in a background plasma density on the order of $10^{19}$~cm$^{-3}$.
	
	A key advantage of using photon acceleration to generate XUV pulses is that the acceleration process preserves the structure of the trailing pulse. As a result, linear optical techniques based on conventional, adaptive, or metasurface optics, can be used to prepare the initial structure of an optical pulse that is injected into the accelerator. The optical pulses used here correspond to the output of a commercially available Ti:sapphire laser system and have a vacuum wavelength $\lambda = 800$~nm, spot size $w = 5.5$~$\mu$m, full width at half maximum (FWHM) duration of 31~fs, and energy $E = 32$~mJ. The pulses are initialized with different vector vortex profiles, which feature a polarization vector that varies in the plane perpendicular to the propagation axis. The profiles are constructed by superposing Laguerre--Gaussian modes in distinct polarization states~\cite{Liu2017GenerationSphere}. Here, the orbital angular momentum value $\ell$ is changed while only the zeroth-order radial mode is used. The vector vortex structure can then be characterized using the bra--ket notation $|\ell,P\rangle$, where $P$ can take the values $V$, $H$, $R$, or $L$ for vertical, horizontal, right circular, or left circular polarization, respectively.
	
	\begin{figure*}[h]
		\centering
		\includegraphics[width=0.98\textwidth]{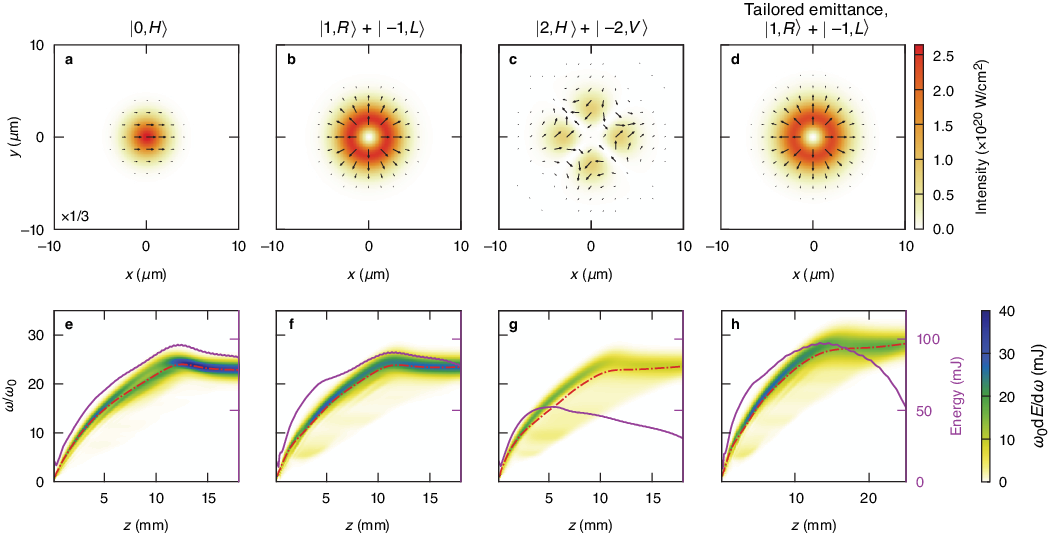}
		\caption{Spatio-polarization profiles and spectral evolution of vector vortex pulses accelerated in an electron beam--driven plasma wave. (a--d)~The transverse intensity and polarization profile near the end of the accelerator for the temporal slice with the maximum intensity. (e--h)~The spectral energy density along the length of the accelerator. An electron beam with uniform emittance drives a density modulation that photon accelerates (a,e)~a Gaussian pulse with linear polarization, (b,f)~a radially polarized pulse, and (c,g)~a pulse with ``spiderweb'' polarization. In each case, the initial vector vortex structure is preserved, and the average frequency upshifts by more than a factor of 20. (d,h)~When the slice emittance of the electron beam is structured to compensate the transverse forces of the plasma wave, the radially polarized pulse upshifts by a factor of 26.  In the bottom row, the red dashed--dotted line is the average frequency and the purple line is the pulse energy within the first period of the plasma wave behind the electron beam. The intensity in (a) is scaled down by a factor of 3 for visualization.}\label{fig:spectrum}
	\end{figure*}
	
	Figure~\ref{fig:spectrum} displays the results of quasi-3D PIC simulations performed with \textsc{Osiris} that demonstrate photon acceleration of vector vortex pulses from the optical to the XUV (see Methods for simulation details). Three equal-energy vector vortex pulses were considered: (a,e)~a linearly polarized Gaussian pulse $\ket{0, H}$ with an intensity $I = 2.1 \times 10^{18} 
	\; \mathrm{W/cm^2}$; (b,f)~a radially polarized pulse $\ket{1, R} + \ket{-1, L}$ with $I = 7.9 \times 10^{17} 
	\; \mathrm{W/cm^2}$; and (c,g)~a pulse that has ``spiderweb'' polarization $\ket{2, H} + \ket{-2, V}$ with $I = 5.8 \times 10^{17}
	\; \mathrm{W/cm^2}$. The top row shows their transverse polarization and intensity profiles near the end of the accelerator at the time slice where the intensity is the highest. The bottom row shows the evolution of their spectra and energies along the length of the accelerator. In all cases, the pulses maintain their initial vector vortex structure, increase in average frequency (red dashed--dotted line) by more than a factor of 20, reach up to 370 times their initial intensity, and maintain their initial amplitude $a_0$ to within a factor of 2, where $a_0 = 8.5\times10^{-10} [I(\mathrm{W/cm}^2)]^{1/2} \lambda(\mu \mathrm{m})$.
	
	\subsection*{Quality of the photon-accelerated pulses}\label{sec:quality}
	
	\begin{figure*}[h]
		\centering
		\includegraphics[width=0.98\textwidth]{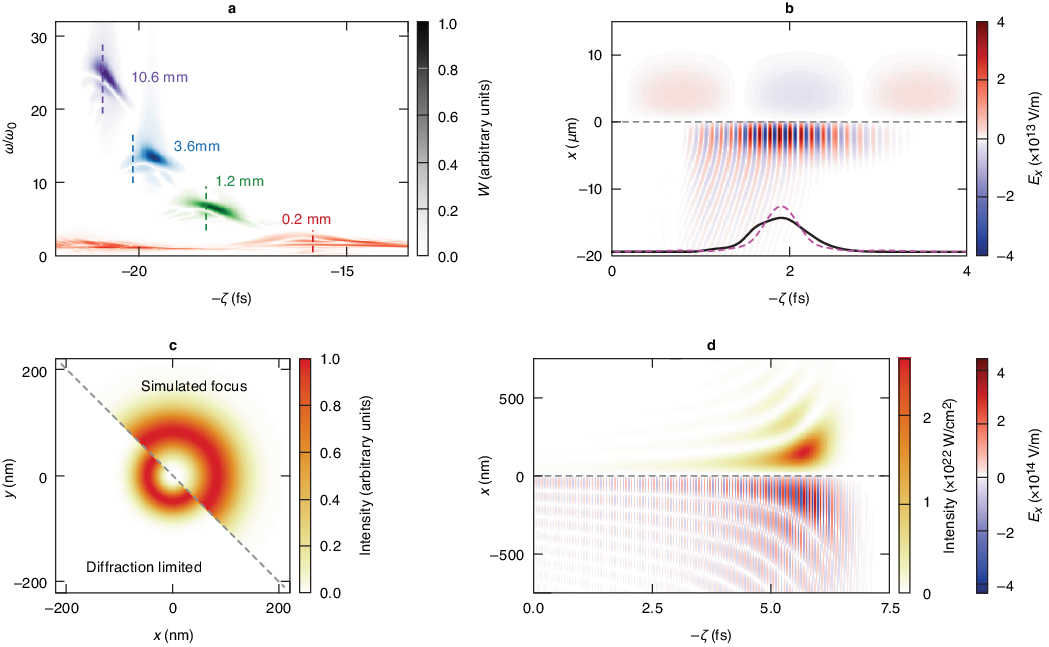}
		\caption{Quality of a radially polarized XUV pulse produced by photon acceleration in a beam-driven plasma wave. The pulse is the same as that presented in Figs.~\ref{fig:spectrum}(b,f), where the electron beam had uniform emittance.  (a)~Wigner distributions $W$ at four $z$ positions along the accelerator, showing the frequency upshift, increase in bandwidth, and temporal compression of the pulse. The dashed lines mark the location of the density spike.  (b)~The electric field of the pulse at the beginning (top) and end (bottom) of the accelerator. The phase fronts are nearly flat, and the final intensity profile (black solid line) is nearly transform-limited (purple dashed line). (c)~The transverse intensity profile of the pulse (top right)~when focused by an ideal $f/3$ lens compared to a  diffraction-limited profile (bottom left). The spot size is only 75\% larger than the diffraction limit. (d)~The intensity (top) and electric field (bottom) of the pulse when focused by an $f/3$ plasma lens.}\label{fig:snapshots}
	\end{figure*}
	
	Photon acceleration converts the optical, radially polarized pulse to an ultrashort, high-intensity, structured XUV pulse with a central wavelength of 36~nm, intensity $I = 2.5\times 10^{20}$~W/cm$^2$, near-transform-limited FWHM duration of 710~as, and flat phase fronts. Figure~\ref{fig:snapshots} examines the evolution and properties of this pulse [Figs.~\ref{fig:spectrum}(b,f)] in greater detail. The magnitude of the Wigner distributions $W$ for the pulse are displayed in Fig.~\ref{fig:snapshots}(a), revealing that the pulse is initially divided by the density spike. The portion in front of the spike accelerates and compresses temporally. The compression can be observed in the phase-space rotation of the Wigner distribution from a temporally (horizontally) extended structure to a spectrally (vertically) extended structure.
	
	Figure~\ref{fig:snapshots}(b) shows that the frequency-upshifted pulse has flat phase fronts and a near-transform-limited duration. The top and bottom panels compare the electric fields of the pulse initially and near the end of the accelerator, respectively. The temporal compression of both the phase fronts and overall duration are apparent. The temporal profile of the XUV pulse at its peak intensity (black solid line) is close to the transform-limited profile (purple dashed line). The transform-limited profile was obtained by Fourier transforming the field to the frequency domain, setting the spectral phase to zero, and transforming back to the time domain.
	
	To demonstrate the quality of the phase fronts, Fig.~\ref{fig:snapshots}(c) compares the transverse profile of the XUV pulse in the focal plane of an $f/3$ lens (upper right) to the diffraction-limited profile of an ideal radially polarized beam (bottom left). The accelerated XUV pulse has a spot size of 120~nm, which is only 75\% larger than the diffraction-limited spot. The focused profile of the XUV pulse was obtained by applying the phase of an ideal $f/3$ lens and then simulating the propagation to the focal plane using the unidirectional pulse propagation equation (UPPE)~\cite{Kolesik2002UnidirectionalEquation}.
	
	In practice, achromatic focusing of such intense and broadband XUV pulses to near-diffraction-limited spots presents a technological challenge. The existing approaches to focusing short-wavelength light, such as toroidal mirrors or Fresnel zone plates \cite{Poletto2013Micro-focusingMirrors,Kubec2022AnLens,Sanli2023ApochromaticFocusing}, cannot withstand high intensities or achieve diffraction-limited spots at small $f$ numbers. As an alternative, the XUV pulses could be focused by a plasma lens~\cite{Palastro2015PlasmaPulses,Li2024SpatiotemporalLens}. Plasma lenses have the advantages of being resistant to high intensities, and thus can be placed in the far field, and are replaceable on a shot-to-shot basis. The downside is that they are inherently chromatic. UPPE simulations show that an $f/3$ plasma lens could focus a radially polarized XUV pulse produced by the photon accelerator to a 200-nm spot size, 800-as FWHM duration, and intensity $I = 2.6\times 10^{22}$~W/cm$^2$ [Fig.~\ref{fig:snapshots}(d)]. If an achromatic focusing scheme could be devised, the focused pulse would have a 120-nm spot size, 680-as duration, and intensity $I = 1.3\times 10^{23}$~W/cm$^2$.
	
\begin{figure*}[h]
		\centering
		\includegraphics[width=0.49\textwidth]{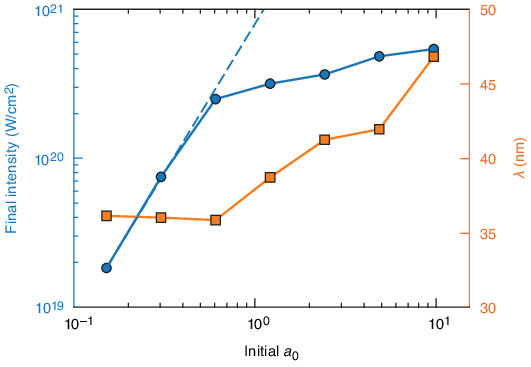}
		\caption{Robustness of the photon accelerator to relativistically intense pulses. The final maximum intensity and average wavelength of the radially polarized pulse near the end of the accelerator as a function of the initial $a_0$. The dashed line shows the ideal intensity scaling, $I \propto a_0^2$. The beam-driven plasma wave is resilient to the relativistic amplitude of the accelerated pulse up to $a_0 \approx 0.6$.}\label{fig:a0-scan}
	\end{figure*}

	Both the initial optical pulse and final XUV pulse have a relativistic amplitude $a_0 \approx 0.6$. Despite this high amplitude and regardless of the frequency, the pulse does not appreciably modify the density gradient responsible for the acceleration, i.e., the photon acceleration process is robust to high-intensity radiation. When the initial amplitude of the optical pulse is increased beyond $a_0 \approx 0.6$, however, the pulse modifies the electron density in the spike, which disrupts the acceleration process. Figure~\ref{fig:a0-scan} shows how the final average wavelength and intensity of the radially polarized XUV pulse depend on the initial $a_0$. Ideal intensity scaling (dashed line) persists up to $a_0 \approx 0.6$, after which the final wavelength begins to increase and the intensity is limited. Even at $a_0 = 10$, the frequency upshift is greater than a factor of 15.
	
	\subsection*{Evolution of the drive beam}\label{sec:beam}
	
	The maximum frequency upshift of the laser pulse is limited by the evolution of the electron beam. The background electrons expelled by the beam and the surplus of ions that remain generate transverse and longitudinal fields that increase from the head of the beam to the tail. As a result, the back of the beam experiences the largest focusing and decelerating forces. The focusing, in particular, causes the back of the beam to pinch. This changes the location where the density modulation is zero $\zeta_\delta$, so that it no longer tracks the temporal centroid $\zeta_\mathrm{c}$ of the trailing pulse.
	
	Over the width of the beam, the focusing force responsible for the pinching $F_r$ can be approximated as $F_r \approx -\tfrac{1}{2}\alpha(\zeta,z) m_\mathrm{e} \omega_\mathrm{p0}^2 r$, where $r$ is the radial coordinate, $\omega_\mathrm{p0} = [e^2 n_0(0) / m_\mathrm{e} \varepsilon_0]^{1/2}$ is the plasma frequency at $z=0$, and the coefficient $\alpha(\zeta,z)$ depends on the background plasma and beam parameters. The spot size of the electron beam $\sigma(\zeta,z)$ evolves in response to this force as follows~\cite{Lee2019AcceleratorPhysics}:
	\begin{equation} \label{eq:beam}
		\frac{\mathrm{d}^2 \sigma}{\mathrm{d}z^2} + \frac{1}{\gamma} \frac{\mathrm{d}\gamma}{\mathrm{d}z} \frac{\mathrm{d}\sigma}{\mathrm{d}z} = \sigma \left[ \frac{\epsilon_\mathrm{n}^2(\zeta)}{\gamma^2(\zeta,z)\sigma^4} - k_\beta^2( \zeta,z) \right],
	\end{equation}
	where $k_\beta(\zeta,z) = \sqrt{\alpha(\zeta,z) / 2\gamma(\zeta,z)} \omega_\mathrm{p0}/ c$ and $\epsilon_\mathrm{n}(\zeta)$ is the normalized emittance. If the plasma density and beam energy are constant, the beam will propagate with a constant (matched) spot size when the focusing force compensates the natural expansion of the beam, i.e., $k_\beta = \epsilon_\mathrm{n} / \gamma \sigma^2$. For the conditions considered here, however, $F_r$ does not vary linearly with $r$ across the entire beam, the background density $n_0$ depends on $z$, and the electron energy $\gamma$ depends on both $\zeta$ and $z$. This makes it difficult to structure the emittance so that exact matching is satisfied everywhere.
	
	Approximate matching can be achieved by choosing the emittance to match the average focusing force, i.e., $\epsilon_\mathrm{n} = \gamma k_\beta \sigma^2$. The uniform-emittance electron beam used in the simulations had a normalized emittance determined by the average force: $\epsilon_\mathrm{n} = (\gamma_0\bar{\alpha}/2)^{1/2}\sigma_0^2 \omega_\mathrm{p0}/ c =$ 1.2~mm-rad, where $\sigma_0$ is the initial beam radius, $\gamma_0$ is the initial electron energy, and $\bar{\alpha}$ is the average of $\alpha$ over density and $\zeta$ (see Methods for details). Simulations of the beam propagating through the full density ramp shown in Fig.~\ref{fig:schematic}(d) confirmed that this value of emittance slowed the evolution of the beam and optimized the length of the photon accelerator.
	
	An emittance profile that is tailored in $\zeta$ can further slow the beam evolution, enabling higher-intensity and shorter-wavelength XUV pulses. The tailored-emittance electron beam used for Figs.~\ref{fig:spectrum}(d,h) had a slice-dependent normalized emittance given by $\epsilon_\mathrm{n}(\zeta) = [\gamma_0\tilde{\alpha}(\zeta)/2]^{1/2}\sigma_0^2 \omega_\mathrm{p0}/ c$, where $\tilde{\alpha}(\zeta)$ is the average of $\alpha$ over density (see Methods for details). Comparing Figs.~\ref{fig:spectrum}(d,h) and (b,f) shows that the tailored-emittance profile increases the energy and frequency of the XUV pulse by $12\%$ (final average wavelength $\lambda = 32$ nm).
	
	\section*{Discussion}\label{sec:conclusion}
	
	A relativistic electron beam can drive a plasma wave that photon accelerates high-intensity vector vortex laser pulses from the optical to the XUV over a few centimeters. The efficiency of the frequency conversion can be defined in several ways. In the case of the radially polarized pulse, the energy of the electron beam decreased from 290 to 280~J, while the pulse energy in the first period of the plasma wave increased from 14 to 87~mJ. This transfer of energy from the beam to the pulse corresponds to an efficiency of $7.3\times 10^{-3}$, which compares favorably to other schemes for generating XUV light~\cite{Dromey2006HighLimit,Versolato2022Microdroplet-tinReview}.
	
	The frequency conversion in a plasma-wave--based photon accelerator conserves the number of photons and phases~\cite{Mora1997,Mori1997,Mendonca2000TheoryAcceleration}. The conservation of photon number can be used to determine (approximately) the fraction of initial photons that are trapped and accelerated by the plasma wave, or ``trapping efficiency.'' The number of photons $\mathcal{N} \propto \omega a_0^2$ and energy $E \propto \omega^2 a_0^2$ are related by $\mathcal{N} =  E/\omega$. If the entirety of the initial pulse contained in the first plasma period were accelerated, the final (f) energy would exceed the initial (i) energy by the ratio of the final and initial frequencies: $E_\mathrm{f} = (\omega_\mathrm{f}/\omega_\mathrm{i})E_\mathrm{i}$. As observed in Fig.~\ref{fig:spectrum}, however, the final energies of the XUV pulses are somewhat lower: $E_\mathrm{f}/E_\mathrm{i} \approx 6$, while $\omega_\mathrm{f}/\omega_\mathrm{i} \approx 22$. This is because a portion of the photons eligible for acceleration in the first period of the plasma wave escape laterally due to diffraction. This effect is most pronounced for the ``spiderweb'' pulse, whose radial extent is comparable to the transverse width of the density spike [Figs.~\ref{fig:spectrum}(c,g)]. Evaluating the right-hand side of the relation $\mathcal{N}_\mathrm{f}/\mathcal{N}_\mathrm{i} = (\omega_\mathrm{i}/\omega_\mathrm{f})(E_\mathrm{f}/E_\mathrm{i})$ with the results for the radially polarized pulse indicates that approximately 30\% of the initially eligible photons were trapped and accelerated. 
	
	The background plasma density $n_0(z)$ was tailored to optimize the frequency-upshift of a pulse trapped in the first period of the plasma wave [Eqs.~\eqref{eq:freq} and \eqref{eq:den}]. Because the initial pulse spanned multiple periods [see Fig.~\ref{fig:schematic}(a)], a substantial portion of the pulse was also trapped in the second period. 
	Despite the suboptimal density profile for acceleration in the second period, this portion upshifted to the XUV. For the radially polarized pulse pictured in Figs.~\ref{fig:spectrum}(b,f), the second pulse reached an average wavelength $\lambda = $ 51~nm, energy $E=$ 120~mJ, and intensity $I = 2.7\times 10^{20}$~W/cm$^2$. In principle, an initial pulse with a longer duration that overlapped many periods could produce a train of XUV pulses separated by ${\approx}$20 fs, but with gradually increasing wavelengths. Such a train may have utility as a sequence of probes for ultrafast, time-resolved measurements.
	
	The design of the photon accelerator was motivated by electron-beam parameters previously achieved at SLAC and laser pulses produced by commercially available Ti:sapphire amplifiers~\cite{Danson2015PetawattWorldwide}. The 50-GeV electron beam evolved minimally over 1.2~cm: the electrons in the tail of the beam lost at most 4~GeV, or 8\% of their initial energy. This suggests that lower-energy drive beams with ${>}4$-GeV electrons could be used to obtain comparable frequency shifts. However, mitigating beam evolution at lower energies may require a slice-by-slice transverse profile and emittance as determined by a more-complete theoretical model, e.g., a generalization of Eq.~\eqref{eq:beam} to higher radial moments. Optimal structuring of drive beams is an outstanding area of research in beam dynamics that also has applications to collider physics~\cite{Zhao2020}. Such an investigation is outside the scope of this work, but provides a starting point for future studies. 
	
	Further optimization of the photon accelerator could be achieved by tuning the plasma, electron beam, or laser pulse parameters. The photon accelerator inherently provides a tunable-wavelength source: the length of the plasma can be adjusted to obtain any frequency between the initial and final frequencies. A wider electron beam with the same charge density would drive a plasma wave with a broader density spike, allowing for the acceleration of laser pulses with larger spot sizes and more energy. In principle, the initial temporal profile or phase of the laser pulse could be structured to control the final bandwidth or duration, either through linear or nonlinear interactions with the plasma wave. As a final example, simulations of vector vortex pulses initialized with $\lambda = 400$~nm (i.e., frequency-doubled Ti:sapphire pulses) yielded an XUV pulse with $4\times$ higher peak intensity ($1.5 \times 10^{21}$~W/cm$^2$), comparable final wavelength (47~nm), and comparable quality as pulses initialized with $\lambda = 800$~nm accelerated over the same distance. The higher intensity resulted from better trapping efficiency; the $f/N$ was the same between the two cases, but longer-wavelength pulses are more susceptible to plasma refraction and lateral losses due to their larger spot size.
	
	In summary, a nonlinear plasma wave driven by a relativistic electron beam can frequency upshift optical vector vortex pulses to the XUV while preserving the initial vector vortex structure. Boosted-frame, quasi-3D PIC simulations demonstrated the formation of relativistically intense  ($10^{20}$~W/cm$^2$), attosecond (710~as), vector vortex XUV pulses with $\lambda = 36$-nm average wavelengths---frequency upshifts by more than a factor of 20. The pulses are nearly transform-limited in duration and have flat phase fronts that can be focused to within 75\% of the diffraction-limited spot size. The photon acceleration process is largely independent of laser polarization and is mainly affected by the radial extent of the pulse, indicating that, in principle, any vector vortex profile could be accelerated to the XUV. The tunable XUV source proposed here bridges the gap in wavelength, structure, and intensity between traditional lasers and existing XUV light sources, providing new tools for scientific discovery.

	\backmatter
	
	
	\section*{Methods} \label{sec:methods}
	
	\subsection*{Tailored density profile}
	Tailoring the longitudinal profile of the plasma density ensures that the accelerating pulse remains colocated with the density gradient of the plasma wave~\cite{Sandberg2023PhotonXUV}. The density profile $n_0(z)$ for which the locations of the temporal centroid of the pulse $\zeta_\mathrm{c}$ and zero density modulation $\zeta_\delta$ are equal can be found by solving the coupled system of Eqs.~\eqref{eq:freq} and \eqref{eq:den}. In order to close this system, analytic or numerical solutions are needed for $\zeta_\delta(n_0)$ and the density gradient $\partial_\zeta \delta n(\zeta,n_0)|_{\zeta = \zeta_\delta}$ as functions of the background density. Here, these solutions are obtained numerically using a 1D, quasistatic model \cite{Rosenzweig1987NonlinearAccelerator, Sandberg2024DephasinglessAcceleration}. In this model, an electron beam with density $n_\mathrm{b}(\zeta)$ drives a plasma wave whose normalized electrostatic potential $\phi(\zeta,n_0) = e\Phi/m_\mathrm{e} c^2$ evolves according to
	\begin{equation}
		\frac{\partial^2 \phi}{\partial \zeta^2} = \frac{1}{2}\omega_\mathrm{p}^2\left[\frac{1}{(1+\phi)^2} - 1 + \frac{2n_\mathrm{b}(\zeta)}{n_0} \right],
	\end{equation}
	where $\omega_\mathrm{p} = (e^2 n_0 / m_\mathrm{e} \varepsilon_0)^{1/2}$. Upon solving this equation for $\phi(\zeta,n_0)$, the electron density modulation can be calculated as
	\begin{equation} \label{eq:dn}
		\delta n(\zeta,n_0) = \frac{n_0}{2}\left\{ \frac{1}{\left[1+\phi(\zeta,n_0)\right]^2} - 1\right\}.
	\end{equation}
	This expression can then be used to find $\zeta_\delta(n_0)$ and $\partial_\zeta \delta n(\zeta,n_0)|_{\zeta = \zeta_\delta}$. Based on the results of the PIC simulations, the 1D quasistatic model does an excellent job of approximating the ideal background density profile. In an experiment, production of this continuous density profile may present a challenge. However, as demonstrated in Ref.~\cite{Sandberg2023Electron-beam-drivenPhotons}, comparable results can be obtained by using segments of constant-density plasma.
	
	Within the 1D quasistatic approximation the electron density $n(\zeta,z) = n_0(z) + \delta n(\zeta,z)$ can be expressed in terms of the electrostatic potential and electron energy~\cite{Sprangle1990NonlinearPlasmas}:
	\begin{equation} \label{eq:den-pot}
		n(\zeta,z) = \frac{\gamma(\zeta,z)}{1+\phi(\zeta,z)}n_0(z).
	\end{equation}
	As a result, the rate of frequency upshifting depends on the gradient of the potential:
	\begin{equation}
		\frac{\mathrm{d}\omega}{\mathrm{d}z} = \frac{e^2}{2\varepsilon_0 m_\mathrm{e}c^2k_z} \frac{\partial}{\partial \zeta}\left( \frac{1}{1+\phi} \right).
	\end{equation}
	The gradient of the potential $\partial_\zeta \phi$ has an extremum where $\delta n = 0$ [Eq.~\eqref{eq:dn}]. Thus, in 1D, the location where $\delta n = 0$ corresponds to the location where the rate of frequency upshifting is the largest.

	\subsection*{Particle-in-cell simulations}
	
	The bulk of the simulations presented in this work were performed using the PIC code \textsc{Osiris}~\cite{Fonseca2002a}, which solves the fully relativistic equations of motion for particles and employs a finite-difference time-domain solver for the electromagnetic fields. The simulations were discretized in the quasi-3D geometry, which represents quantities as a truncated expansion in azimuthal modes~\cite{Davidson2015}. This geometry is appropriate for systems with a high degree of cylindrical symmetry. The small cell sizes required to properly resolve the upshifted light make lab-frame simulations of the 2-cm-long accelerators prohibitively expensive, so simulations were performed in a Lorentz-boosted frame~\cite{Yu2016,Vay2007NoninvarianceInteractions,Fonseca2008} with $\gamma = 15$. Simulations in the boosted frame are susceptible to the numerical Cerenkov instability (NCI), which can cause spurious radiation to grow from noise. This was mitigated by using a customized solver~\cite{Li2021}, which was found to suppress NCI while simultaneously correcting for numerical dispersion and time-staggering errors in the Lorentz force. Results were interpolated from the boosted frame to the lab frame when plotting the results.
	
	The simulations used a moving window in the boosted frame of size 458~$\mu\mathrm{m}\times76.4$~$\mu$m ($14896\times 1200$ cells) in $z$ and $r$, respectively.  The time step was 5.73~as, and the total simulation duration was 2.91~ps.  Electrons (ions) were simulated with 16--48 (8--16) particles per cell, depending on the number of azimuthal modes used in the quasi-3D geometry.  The number of modes was one larger than the highest $\ell$ mode of each vector vortex pulse. Truncating the number of azimuthal modes can result in inaccurate modeling of 3D effects, such as hosing instabilities. To verify that the number of modes was sufficient, convergence tests were performed by doubling the number of modes for the Gaussian pulse. Compared to the nominal number of modes, the central frequency and intensity of the final XUV pulse only changed by a few percent.
	
	Each laser pulse had an initial wavelength $\lambda = $ 800~nm, spot size $w=$ 5.5~$\mu$m, energy $E=$ 32~mJ, and FWHM duration of 31~fs. The electron beam was initialized with a 0.64-$\mu$m length, $\sigma_{0} = 13$-$\mu$m $e^{-1/2}$ radius, 5.8~nC of total charge, and an average electron energy of $\gamma_0=$ 50~GeV. For the uniform-emittance cases, a normalized emittance $\epsilon_\mathrm{n}=$ 1.2~mm-rad was used; for the tailored-emittance case, the normalized emittance varied linearly from 0.11~mm-rad at the head of the beam to 1.9~mm-rad at the tail.  The background plasma density shown in Fig.~\ref{fig:schematic}(d) with $n_0(0) = 7\times 10^{19}$~cm$^{-3}$ was used in all cases.
	
	To determine the optimal emittance of the electron beam, simulations were performed using \textsc{Qpad}~\cite{Li2021AQPAD}---a quasi-static PIC code that employs azimuthal decomposition. The focused beam was propagated a short distance into plasmas with constant densities ranging from $n_\mathrm{e}$ = 3 to $7 \times 10^{19}$~cm$^{-3}$. At each density, the transverse force of the plasma wave on the beam was fit to $F_r = -\tfrac{1}{2} \alpha(\zeta,n_0) m_\mathrm{e} \omega_\mathrm{p0}^2  r$, where $\omega_\mathrm{p0}$ is the plasma frequency for the maximum density $n_0(0) = 7 \times 10^{19}$~cm$^{-3}$, and $\alpha(\zeta,n_0)$ is a fitting parameter that depends on $\zeta$ and density. For linear plasma waves driven by a Gaussian beam, this parameter can be approximated as $\alpha \approx 2 c^2 \delta n(\zeta, n_0)/n_0 \omega_\mathrm{p0}^2 \sigma^2$~\cite{Chen1987PlasmaBeams}. For the uniform-emittance cases, the emittance was calculated using the density- and $\zeta$-averaged value $\bar{\alpha} = 4.8\times 10^{-4}$ [$\epsilon_\mathrm{n}=(\gamma_0 \bar{\alpha}/2)^{1/2} \sigma_{0}^2 \omega_{\mathrm{p}0}/c$]. For the tailored-emittance case, the slice emittance was calculated using the density-averaged values $\tilde{\alpha}(\zeta)$, which increased from $0$ at the head of the beam to $7\times 10^{-4}$ at the tail.
	
	\bibliography{references-custom}

	\section*{Data availability}
	The datasets used during the current work are available from the corresponding author upon reasonable request.
	
	\section*{Code availability}
	All codes used during the current work are available from the corresponding author upon reasonable request.
	
	\section*{Acknowledgments}
	The authors would like to acknowledge N. Beri, Q. Qian, D.H. Froula, D. Li, and J.L. Shaw for insightful conversations.
	
	This report was prepared as an account of work sponsored by an agency of the U.S. Government. Neither the U.S. Government nor any agency thereof, nor any of their employees, makes any warranty, express or implied, or assumes any legal liability or responsibility for the accuracy, completeness, or usefulness of any information, apparatus, product, or process disclosed, or represents that its use would not infringe privately owned rights. Reference herein to any specific commercial product, process, or service by trade name, trademark, manufacturer, or otherwise does not necessarily constitute or imply its endorsement, recommendation, or favoring by the U.S. Government or any agency thereof. The views and opinions of authors expressed herein do not necessarily state or reflect those of the U.S. Government or any agency thereof.
	
	This material is based upon work supported by the Office of Fusion Energy Sciences under Award Numbers DE-SC0021057, the Department of Energy (DOE) [National Nuclear Security Administration (NNSA)] University of Rochester ``National Inertial Confinement Fusion Program'' under Award Number DE-NA0004144, the DOE High Energy Density Laboratory Plasmas (HEDLP) through the NNSA under Award Number DE-NA0004131, the DOE High Energy Physics under Award Number DE-SC0010064, the Laboratory for Laser Energetics subcontract SUB00000211/GR531765, the National Science Foundation under Award Numbers 2206059, 2108075, and 2108970, and the New York State Energy Research and Development Authority. Simulations were performed at NERSC under m4372.
	
	\section*{Author contributions}
	
	K.G.M. conceived, designed, and conducted the simulations, performed principal analysis, performed the visualizations, and was a principal writer.
	J.P.P. conceived the simulations, performed principal analysis, and was a principal writer.
	A.G.R.T. conceived the simulations, provided analysis of the results, and reviewed the manuscript.
	J.R.P. developed simulation capabilities, provided analysis of the results, and reviewed the manuscript.
	W.B.M. proposed the concept, provided analysis of the results, and reviewed the manuscript.
	F.L. proposed the concept and assisted in simulation design.
	B.K.R. provided analysis of the results and reviewed the manuscript.
	All authors discussed the results.
	
	\section*{Competing interests}
	The authors declare no competing interests.

\end{document}